\documentclass[journal]{IEEEtran}

\usepackage{array} 
\usepackage{epsfig}
\usepackage{epstopdf}
\usepackage{times}
\usepackage{framed}
\usepackage{enumitem}
\usepackage{float}
\usepackage{etoolbox}
\usepackage{algorithmic}
\usepackage{amssymb}
\usepackage{amsmath}
\usepackage{multirow}
\usepackage{url}
\usepackage{lipsum}
\usepackage{soul}
\usepackage{caption}
\usepackage{mathtools}
\usepackage{bbm}
\usepackage{booktabs}
\usepackage{xcolor}
\usepackage{makecell}
\usepackage{pifont}
\usepackage{subfigure}
\usepackage[export]{adjustbox} 
\usepackage[norelsize,boxruled]{algorithm2e}
\usepackage{latexsym}
\usepackage{cite}
\usepackage{multicol}
\usepackage{eurosym}
\usepackage{xspace}
\usepackage{pdfpages}
\usepackage{verbatim}
\usepackage{stfloats}

\makeatletter
\newcommand{\removelatexerror} {\let\@latex@error\@gobble}
\makeatother

\def\1{\mathbf{1}}

\newcounter{subeqn} %
\makeatletter
\@addtoreset{subeqn}{equation}
\makeatother

\newcolumntype{x}[1]{%
>{\centering\hspace{0pt}}p{#1}}%

\usepackage{tikz}
\usepackage{textcomp}
\usepackage{hyperref}
\usepackage{lipsum}
\newcommand\copyrighttext{%
  \footnotesize \textcopyright 2021 IEEE. Personal use of this material is permitted.
  Permission from IEEE must be obtained for all other uses, in any current or future 
  media, including reprinting/republishing this material for advertising or promotional 
  purposes, creating new collective works, for resale or redistribution to servers or 
  lists, or reuse of any copyrighted component of this work in other works. 
  }
\newcommand\copyrightnotice{%
\begin{tikzpicture}[remember picture,overlay]
\node[anchor=south,yshift=10pt] at (current page.south) {\fbox{\parbox{\dimexpr\textwidth-\fboxsep-\fboxrule\relax}{\copyrighttext}}};
\end{tikzpicture}%
}

\begin{document}

\title{First Responders Got Wings: UAVs to the Rescue of Localization Operations in Beyond 5G Systems}
\author{Antonio~Albanese,~\IEEEmembership{Student Member,~IEEE,}
Vincenzo~Sciancalepore,~\IEEEmembership{Senior Member,~IEEE,}
\\ Xavier~Costa-P\'erez,~\IEEEmembership{Senior Member,~IEEE}
\thanks{\textit{A. Albanese is with NEC Laboratories Europe, 69115 Heidelberg, Germany and University Carlos III of Madrid, 28911 Legan\'es, Spain.\newline
V. Sciancalepore is with NEC Laboratories Europe, \newline
X. Costa-P\'erez is with NEC Laboratories Europe, and i2cat Foundation, 08034 Barcelona, Spain and
ICREA, 08010 Barcelona, Spain,\newline
Emails: \{antonio.albanese, vincenzo.sciancalepore, xavier.costa\}@neclab.eu}.\newline
This work was supported by EU H2020 RISE-6G project under Grant 101017011.}%
}
\maketitle
\copyrightnotice

\begin{abstract}
Natural and human-made disasters have dramatically increased during the last decades. Given the strong relationship between \emph{first responders localization time} and the final \emph{number of deaths}, the modernization of search-and-rescue operations has become imperative. 
In this context, Unmanned Aerial Vehicles (UAVs)-based solutions are the most promising candidates to take up on the localization challenge by leveraging on emerging technologies such as: Artificial Intelligence (AI), Reconfigurable Intelligent Surfaces (RIS) and Orthogonal Time Frequency Space (OTFS) modulations.
In this paper, we capitalize on such recently available techniques by shedding light on the main challenges and future opportunities to boost the localization performance of state-of-the-art techniques to give birth to \emph{unprecedentedly effective missing victims localization solutions}.
\end{abstract}

\section{Introduction}

In the past two decades, the number of climate-related disasters has spurred to almost a doubling, affecting more than $4$ billion people and costing around $3$ trillion USD~\cite{DW2020}. When adding the occurrences of human-made disasters or war acts, the need for solutions to cope with emergency situations becomes unquestionable. Promptness of search-and-rescue operations is crucial to reduce the death toll and quickly establishing contact with endangered victims is a priority.

First-responders modus operandi can be roughly labeled as physical, canine and electronic search~\cite{Fema}. Physical search can be performed by non-specialists without requiring complex electronic equipment. However, being based on visual and sound assessments, it is not effective with unconscious victims and poses more risks for rescuers due to the required proximity to possible dangers. Canine search enables identifying unconscious victims under rubble but it is affected by accessibility problems whenever the human scent cannot reach the surface. Electronic search makes an attempt at overcoming some of the above-mentioned operational weaknesses by probing acoustic and vibration signals coming from the conscious victims or by detecting radio signals from rescue beacons purposely worn by the victims (e.g., skiers in an avalanche scenario)~\cite{NASA2015}. 
Nonetheless, achieving ubiquitous accurate localization without pre-deployed ad-hoc sensing devices remains an unsolved challenge.

The increasing penetration rate of User Equipments (UEs) in our society opens up the possibility of using the cellular network to build an \emph{emergency localization system} and support first responders during search-and-rescue operations. Indeed, the upcoming beyond-fifth-generation (B5G) technology by the third generation partnership project (3GPP) is expected to bring in unprecedented advancements across the whole mobile network~\cite{Chowdhury2020}. Although initially it will mostly inherit the benefits achieved by the fifth generation (5G) architecture in the latest stage of its deployment, B5G will pursue ubiquitous mobile ultra-broadband (uMUB) by integrating satellite and Unmanned Aerial Vehicles (UAVs) in its infrastructure, thereby building a 3D network deployment. As in the aftermath of a disaster the ground base stations (BSs) may be severely impaired, the legacy network architecture may fail to effectively support search-and-rescue operations. While satellite-based localization under the umbrella of Global Navigation Satellite Systems (GNSSs) may not be reliable enough for the purpose due to the lack of direct access for network operators through 3GPP-standardized interfaces and their high sensitivity to channel fading in presence of debris or rubble typical of a disaster scenario, UAVs are gaining more and more interest in emergency operations for their capability to operate in difficult-to-reach locations~\cite{Li2019}.  

The usage of UAVs brings up a series of technical challenges when it comes to victims localization. Although in principle UAVs may directly implement classical cellular-based localization techniques, such techniques are developed for static anchor points, i.e. base stations, thereby missing the new opportunities introduced by their motion capabilities. In particular, we show that one moving UAV performing distance measurements from a slow-moving target can achieve high localization accuracy by exploiting both temporal and spatial information of measurements taken along a closed motion trajectory. Our single-UAV solution minimizes complexity and fosters deployment readiness, which is crucial in emergency conditions, while being easily extendable to multi-UAV deployments by limiting the scope of each UAV on a single disjoint target subarea. The latter approach improves the overall localization accuracy as the average distance between one UE and its serving UAV decreases, thus alleviating the associated measurement error.

In conjunction with improved localization techniques, promising technologies can come into play and improve the overall localization accuracy. Orthogonal Time Frequency Space (OTFS) modulation may turn the high Doppler spread induced by a fast moving UAV into an advantage by working in the delay-Doppler domain instead of the canonical time-frequency domain~\cite{Hadani2017}. Besides, Reconfigurable Intelligent Surfaces (RISs) as revolutionary technology
may come to help overcoming the shadowing effect induced by rubble in the scenario. In particular, considering agility and modularity properties in their future design, RIS could become a reliable piece of ground infrastructure even when most of conventional network devices would be heavily impaired in the aftermath of a disaster, thereby creating controllable propagation conditions for the cellular signals employed in the measurement process~\cite{Kisseleff2021}.  

The remainder of this paper is the following. Section~\ref{s:emergency} provides a summary of UAV-based solutions for establishing connectivity in emergency conditions; Section~\ref{s:multilateration} describes the limitations of traditional localization approaches in such scenarios and introduces the idea of pseudo-multilateration; Section~\ref{s:enablers} presents the key-enablers to effectively solve the localization problem, while envisioning the usage of emerging technologies to further improve the performance. Finally, Section~\ref{s:conclusions} concludes the paper.

\begin{figure}
    \centering
    \includegraphics[width=\linewidth]{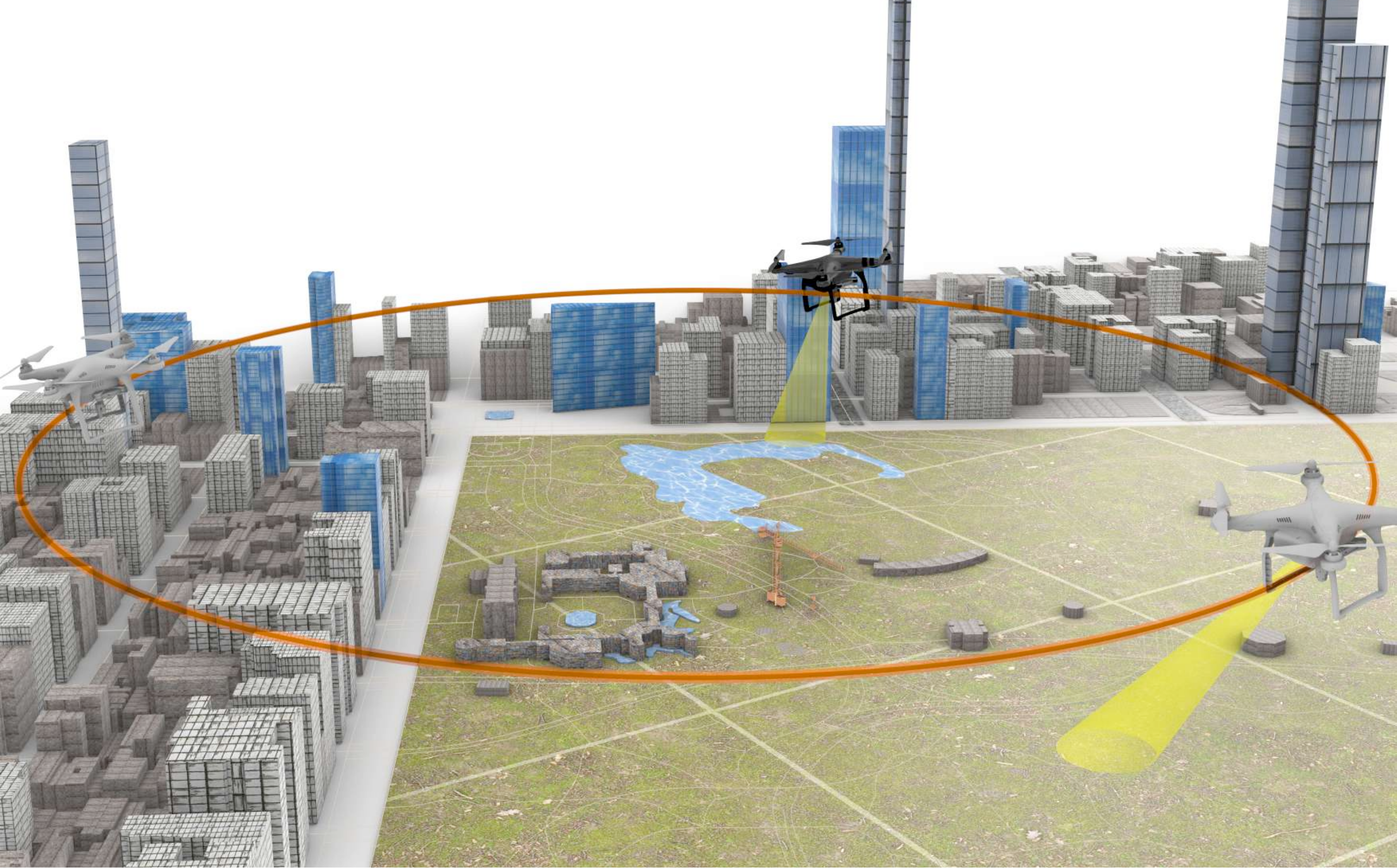}
    \caption{Localization operations via UAV-based Search-and-Rescue in an emergency scenario.}
    \label{fig:arch}
\end{figure}

\section{UAVs for emergency connectivity}
\label{s:emergency}

UAVs flexibility paves the way to their increasing success as a mean to provide connectivity in harsh conditions~\cite{Panda2019}. Unfortunately, such degree of flexibility comes at the cost of high deployment complexity. Indeed, UAVs trajectory planning takes into account stringent flight time constraints due to their limited battery life and payload weight, as well as the target quality-of-service (QoS) at the users side. However, miniaturization of BS technology and greater communication power efficiency allow achieving limited impact on UAVs battery life even when employing commercial off-the-shelf hardware, thereby making UAVs able to operate as portable BSs given appropriate battery replacement during operations. Moreover, when deploying a UAV swarm, collision avoidance is fundamental and requires the solution of non-convex optimization problems to obtain the set of feasible UAV positions at each time instant. 

A UAV-based cellular localization system needs to interact with network operators in order to setup flying base stations (BSs) and perform ranging measurements. In particular, UAV BSs may connect to operators core network and retrieve all the necessary keys to authenticate users and provide cellular coverage. Besides, whenever the network infrastructure is down due to a disaster, operators could provide first-responders with the last-updated record of users in the area (e.g., their International Mobile Subscriber Identity (IMSI)), which, in turn, could be used to improve the effectiveness of search-and-rescue operations. Although workarounds are available, e.g. by means of an UAV acting as a rogue base station implementing an IMSI-cather module~\cite{albanese2021}, B5G emergency localization requires some level of operator support for reliable on-field operations.     
    
Standalone solutions may circumvent the aforementioned dependency on network operators granting access to their own infrastructure. On the one hand, UAVs may employ diverse on-board sensors such as visible, infrared or thermal camera modules to perform victims localization~\cite{Aeryon}. On the other hand, they may use different radio technologies to illuminate the disaster area. For instance, a UAV swarm may setup a two-tier wireless mesh network wherein the lower tier connects the victims via Bluetooth Low Energy (BLE) and the upper tier establishes long-range radio links among the UAVs~\cite{ferranti2019}. 
However, such solutions would require target UEs to execute dedicated services and applications, thus relying on previous users engagement and lowering their effectiveness in emergency conditions. Indeed, accessibility and ubiquity are major requirements for the success of search-and-rescue operations and are likely not achieved by non-cellular-based systems.  

Few Global Navigation Satellite System (GNSS) are able to deliver sub-meter localization accuracy~\cite{Galileo2020}. Though expected to be integrated in the B5G network infrastructure, GNSSs heavily suffer from shadowing introduced by foliage, bodies of water, wooden obstacles and so forth. As reliability is key in ensuring emergency localization, low-altitude UAVs may overcome the power losses introduced by blocking rubble and allow victims UEs to receive useful cellular signals. 
Acting as flying base stations, UAVs may implement traditional localization techniques in the sky. Unfortunately, these techniques are designed for static ground base stations, thereby missing the opportunities introduced by the UAVs mobility.

\section{Single-anchor Multilateration}
\label{s:multilateration}

In the following, we discuss the limits of a direct application of legacy cellular localization techniques to B5G, while introducing a novel approach, namely pseudo-multilateration, tailored to this use case and robust to noise and high-mobility conditions. Although multi-antenna BSs within the context of B5G enable angle-based localization, such techniques require complex algorithms to estimate scatterers positions in the environment and accurately retrieve the UE position. Conversely, ToA-based localization boils down the complexity while offering good accuracy being less sensitive to multipath fading especially at millimeter-wave frequencies, where only few dominant paths (the least scattered) are relevant for communication and localization.   

\subsection{Legacy multilateration}
Multilateration is a classical localization technique for cellular networks based on distance measurements from fixed reference points, namely anchors. Three reference points are sufficient to retrieve 2D target coordinates while at least one more anchor is required in case 3D coordinates are sought. 

Intuitively, the more base stations retrieving distance measurements from a target UE, the higher is the achievable localization accuracy. This result is yielded from the definition of Cram\'er-Rao Lower Bound, which constitutes a lower bound on the localization accuracy for any unbiased localization system and decreases with the number of base stations. Therefore, the cooperation of a high number of anchor base stations is required to jointly measure the instantaneous distances from the target so as to satisfy the demanding requirements in terms of localization accuracy promised by B5G. While an ideally dense deployment of base stations delivers the minimum localization error, this approach would lead to massive overhead due to the processing of several measurement samples.

Distance measurements are based on time-of-arrival (ToA) estimates of both uplink and downlink reference signals (see e.g., Positioning Reference Signals in the current 5G standard). Therefore, as real wireless channels are hindered by fading, multilateration performance is likewise affected. For reference, let us consider a 3D scenario. In ideal channel conditions, multilateration finds the intersection of all the spheres centered at the coordinates of each anchor point having radius equal to the respective measured distance from the target. However, due to channel fading, multiple intersection points between any pair of spheres may be possible, thus leading to an approximated location of the target. Specifically, the target position may be derived by minimizing the sum of the squared distance errors at a non-negligible computational cost. The time complexity of this approach depends on the convergence speed of the selected iterative process. Although fine tuning the algorithm parameters may yield good performance within reasonable time, targets mobility exacerbates the complexity thereby limiting the achievable accuracy of a multilateration localization system.

\subsection{Pseudo-Multilateration}
\label{s:pseudo-mult}
High localization accuracy by means of multilateration requires very dense base stations deployment, either on the ground or in the sky. This requirement stems from the localization system design itself, which assumes static base stations disregarding any ability to move. Traditional multilateration presents the following main drawbacks: $i$) vulnerability to wireless channel slow fading and fast fading, which may alter the distance measurements and consecutively the estimated target location and $ii$) vulnerability to user mobility during the measurement process leading to inconsistent location estimates.

\begin{figure}[t!]
    \centering
    \includegraphics[clip, width=0.85\linewidth]{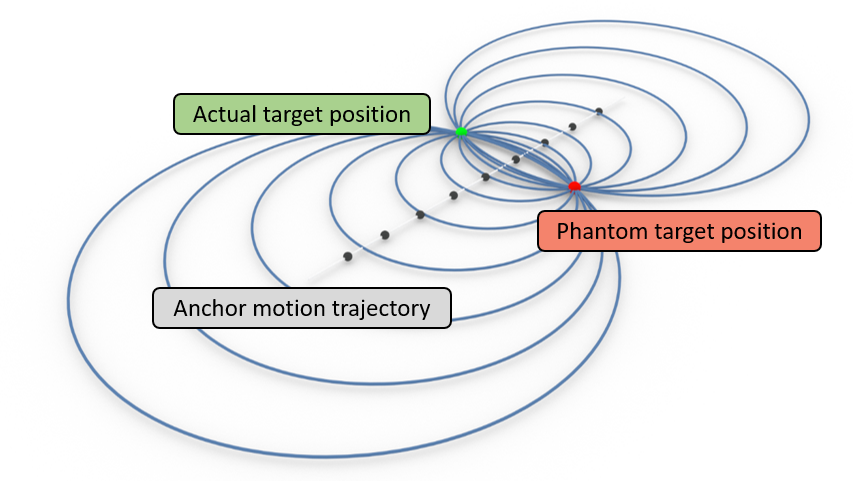}
    \caption{2D localization of a static target with a linear anchor motion trajectory returning a double solution.}
    \label{fig:geom}
\end{figure}

The novel pseudo-multilateration technique overcomes the above-mentioned limitations by reducing the coverage density of anchors and straightforwardly dealing with high-mobility users. First, a single UAV anchor retrieves a set of distance measurements over time along some motion trajectory. Then, such measurements undergo a processing phase whose output is the target trajectory (if moving) within the considered time frame~\cite{albanese2021}. Similar to classical multilateration, pseudo-multilateration objective is to minimize the localization error, i.e. the squared difference between the estimated and the actual distances, at every anchor trajectory point over time. By having only one anchor point, the target position at each time instant is any point lying on a sphere with radius equal to the measured distance and centered on the current UAV position. Since the problem as stated admits infinite solutions, we need to consider a series of measurements performed at several time instants to derive a meaningful estimate. 

The setup of the UAV trajectory is fundamental to successfully perform the localization task. For instance, Fig.~\ref{fig:geom} shows that pseudo-multilateration is not guaranteed to have a unique solution with a linear trajectory but rather returns a double solution, comprising the actual target position and a phantom one. Indeed, we can prove that the UAV must change its direction within a finite time, ideally in a continuous fashion~\cite{albanese2021}. Beside the geometrical considerations, communication constraints force the UAV to hover as close as possible to the victims in order to keep the received signal-to-noise ratio (SNR) above threshold. In particular, the higher the distance, the higher is the measurement error variance as the increased number of indirect propagation paths exacerbates the task of correctly estimating the time-of-arrival of the dominant path at the receive side. 

Circular trajectories serve the purpose while being easy to set up and handle. We would like to point out that although fluctuations of the selected trajectory might happen due to external UAV perturbations, e.g., wind gusts, or obstacles in the way, we assume the position of the UAV to be instantaneously known and accounted for in the problem formulation, as described in Section~\ref{s:ai}. After selecting the right trajectory type, we may write an optimization problem with the objective of minimizing the sum of the Euclidean distances between every target estimate (i.e., solutions pair) obtained at consecutive times. However, such problem is NP-Hard, thus intractable within affordable time. Moreover, it requires the specification of the initial user position from which to build a sequence of distance differences to minimize. Therefore, any suboptimal solution obtained via some heuristics depends on the initialization, often resulting in meaningless estimated users trajectories.  

\begin{figure}[t!]
      \centering
      \includegraphics[clip, width=\linewidth ]{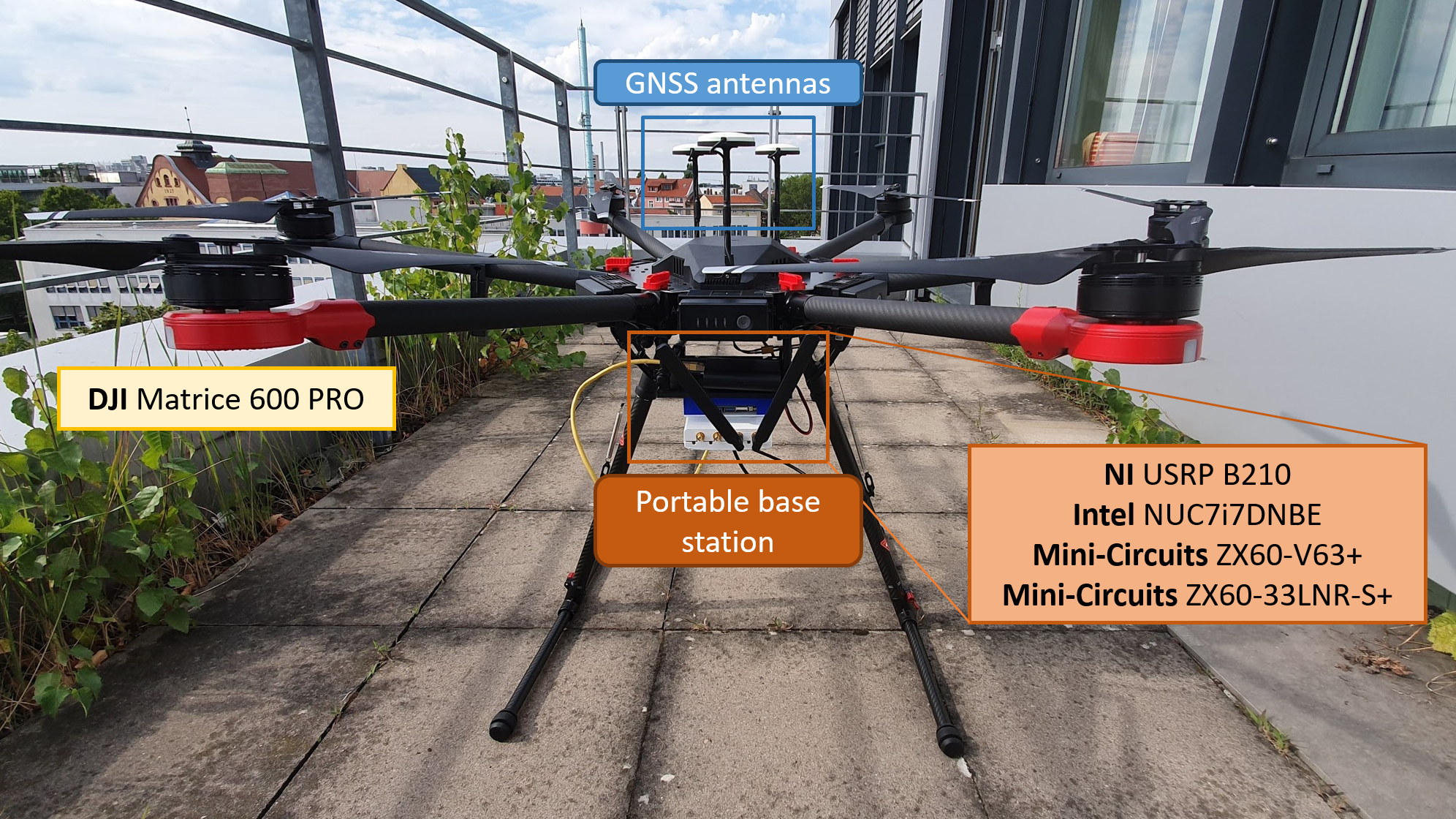}
      \caption{Prototype UAV equipped with a portable base station enabled to perform the pseudo-multilateration technique~\cite{albanese2021}.}
      \label{fig:drone}
\end{figure}

In Fig.~\ref{fig:drone}, we depict a flying UAV equipped with a small cellular base station. Leveraging on the increasing miniaturization of Software Defined Radio (SDR) and system on a chip (SoC) technologies, a powerful commercial off-the-shelf UAV may easily carry and power a portable base station, enabling the execution of the pseudo-multilateration technique in the field.

\section{B5G localization enablers}
\label{s:enablers}
We tackle the B5G localization problem from three diverse and complement perspectives, namely coding, modulation and propagation. Though apparently orthogonal, these three factors contribute to substantial B5G localization accuracy improvements.

\subsection{CNN-based localization}
\label{s:ai}
As the closed-form optimization model mentioned in Section~\ref{s:pseudo-mult} fails to deliver accurate user position estimates, we leverage on machine learning to learn the users
motion behavior and improve localization performance.
The key for accurate user localization lies in the inherent temporal and spatial correlation among the ToA measurements taken by the UAV along its circular trajectory. 
We can preserve such information by arranging the measurements and their corresponding UAV coordinates in matrices so as to build one matrix per set of measurements performed during a single UAV complete revolution.

 \begin{figure}[t!]
    \centering
    \includegraphics[clip, trim={5cm 0 4cm 0}, width=0.95\linewidth]{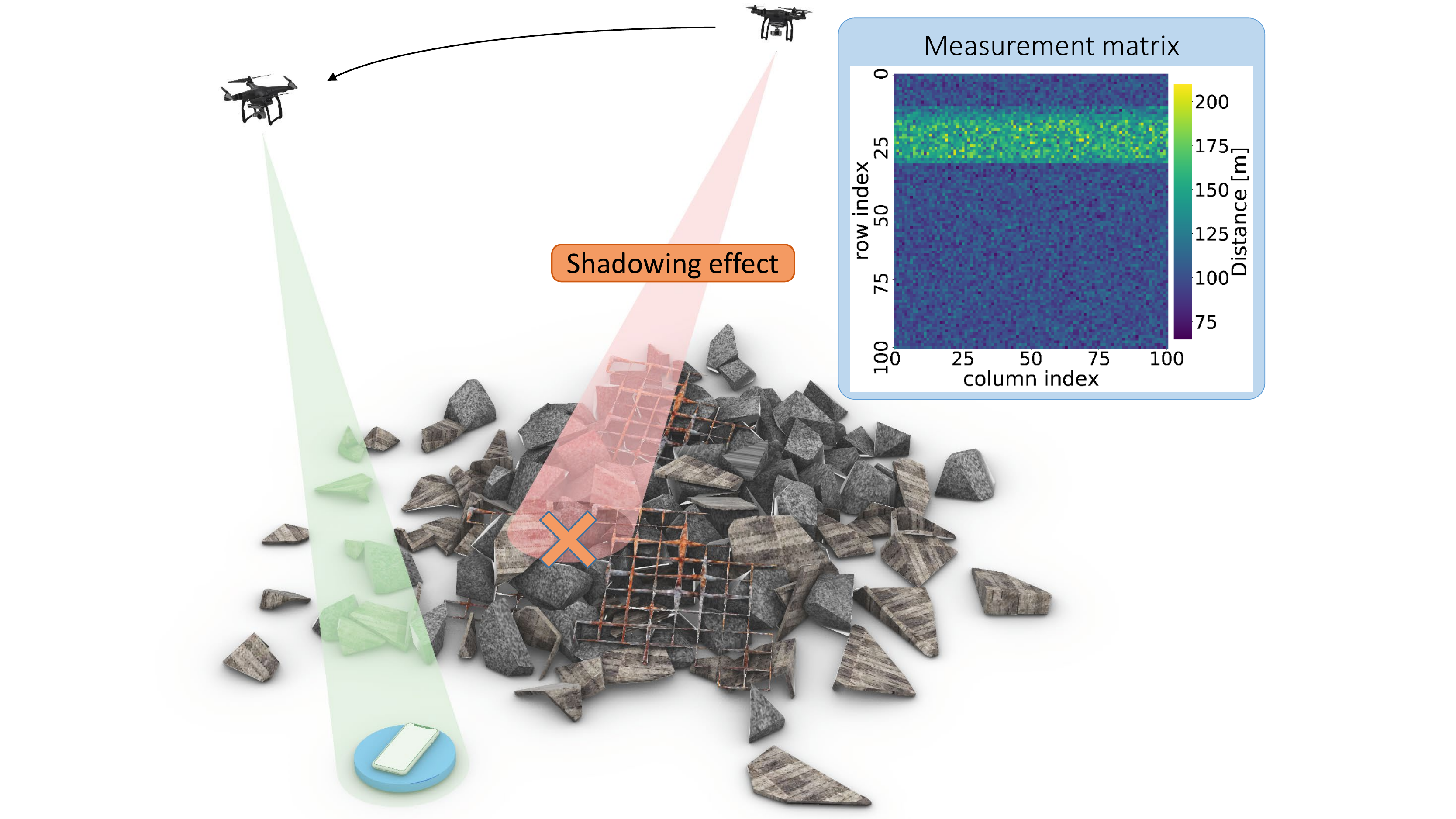}
    \caption{Pseudo-multirateration takes advantage of stripe patterns in the measurement matrix to resolve the channel fading introduced by obstacles in the environment.}
    \label{fig:data_sample}
\end{figure}

A Convolutional Neural Network (CNN) processes such matrix as a single-channel image and outputs the estimated user position coordinates.
Indeed, as shown in Fig.~\ref{fig:data_sample}, any obstacle blocking the line-of-sight (LoS) link between the UAV and the target user during the measurement process would yield a specific stripe pattern in the matrix due to the increased error at the measurements spots in which the wireless channel is affected by the shadowing effect introduced by the obstacle itself.
The CNN can be trained via Supervised Learning over synthetic data obtained by e.g., ray-tracing simulators and Continual Learning (CL) solutions may smoothly update the model to take into account new localization and mobility scenarios while retaining the retrieved useful knowledge over time.     

As already mentioned, circular trajectories ensure that the UAV never gets too far away from the target user, thereby lessening the received SNR to an unusable level. However, even selecting a fixed hovering altitude, circular trajectories require at least two parameters to be completely defined, namely radius and center coordinates, which we proactively compute. Specifically, after every complete UAV revolution, a relocation procedure manoeuvres it onto a new trajectory to move it closer to the target users in order to obtain more accurate distance measurements.      

To this aim, the user location estimates retrieved by means of the aforementioned CNN are fed to a prediction model. Many neural network architectures have proved to be effective in this context. For instance, we may use Long Short-Term Memory (LSTM) cells to build an Encoder-Decoder model and leverage on a fixed-length internal representation of the input sequence to output a sequence of the same or different length, i.e., the prediction of the user position.   

\subsection{Doppler-resilience with OTFS}
B5G will rely on millimeter-wave communications, which are characterized by sparse multipath fading and much larger bandwidth than the classical sub-6~GHz cellular band employed by the current generation. Transmitting at such high frequencies introduce several complications due to increased free-space losses, phase noise and sensitivity to terminal mobility (channel Doppler spread). The first issue can be compensated for by means of antenna arrays providing the required beamforming gain, while the remaining two call for a new modulation technique able to achieve near-constant channel gain even in channels with high Doppler spread.

Indeed, the recently proposed Orthogonal Time Frequency Space (OTFS) modulation is a major waveform contender for B5G being (ideally) independent of the channel Doppler spread, thus resilient in high-mobility scenarios. The OTFS rationale is to expand the basis waveform over the whole time-frequency plane by generating symbols in the delay-Doppler domain and processing them by means of a 2D linear transform dubbed as the Inverse Symplectic Fourier Transform, namely a double Fourier transform with a symplectic inner product in the exponent.  

This approach converts an Orthogonal Frequency-Division Multiplexing (OFDM) fading time-variant channel into a non-fading time-invariant channel~\cite{Hadani2017}. Though its advantages in terms of communication capabilities are apparent~\cite{Raviteja2018}, here we suggest that employing OTFS as B5G physical layer technology may deliver higher localization accuracy with respect to the current 5G-standard OFDM, especially in high-mobility scenarios with low probability of line-of-sight between the BS and the UE.

\begin{figure}[t!]
      \centering
      \includegraphics[clip, width=\linewidth ]{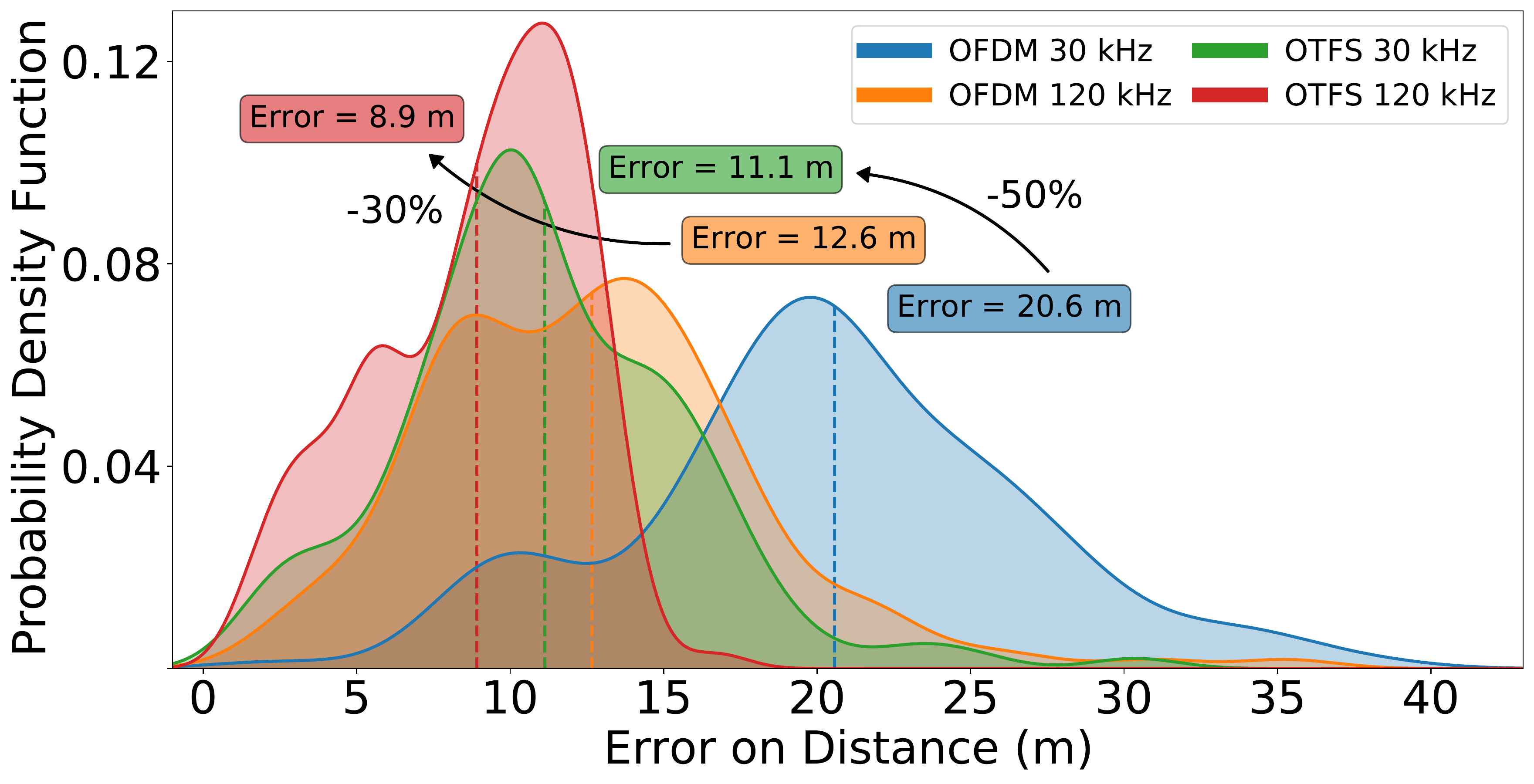}
      \caption{Probability density functions of single-link distance estimation accuracy by means of OFDM and OTFS modulations. In high-mobility scenarios, with little to no line-of-sight conditions, OTFS outperforms OFDM by (up to) $50$\% on average.}
      \label{fig:otfs}
\end{figure}

In order to isolate the performance improvements provided by the modulation scheme, we solely focus on the distance measurement error between a UAV BS and a target UE on the ground. Fig.~\ref{fig:otfs} shows the probability density function (pdf) of the absolute measurement error in non-line-of-sight (NLoS) conditions for a UAV flying at $100$~m altitude and moving along a linear trajectory with speed equal to $10$~m/s. The UAV is equipped with an omnidirectional antenna and employs OTFS or OFDM modulations configured with two different subcarrier spacings, i.e., $30$~kHz and $120$~kHz. The higher the subcarrier spacing, the lower is the system sampling time or, in other words, the better is the resolution of the ToA estimates. As distance measurements are obtained by multiplying ToA estimates by the speed of light, we enjoy finer grained observable distances. 

Increasing the subcarrier spacing diminishes the errors of the distance estimates provided by both modulations in an analogous fashion. However, in this scenario, the OTFS scheme outperforms OFDM by around $50\%$ or $30\%$ (on average) with $30$~kHz or $120$~kHz subcarrier spacings, respectively. Moreover, we note that the OTFS measurements show less variance than the OFDM ones, based on the widths of their pdf peaks, thereby suggesting higher resilience against NLoS propagation conditions and Doppler spread induced by the moving UAV. Indeed, OTFS has a lower probability of returning measurements affected by very high error compared to OFDM in the same challenging scenario.      

\subsection{Location-awareness Propagation Environment}
\label{s:ris}

As previously argued in Section~\ref{s:ai}, properly designed Artificial Intelligence solutions may revert the signal distortions introduced by harsh environment and deliver an estimate of the user position when classical mathematical optimization fails. However, particularly complex environments might still be show-stoppers when trying to accurately localize victims in disaster areas even leveraging on AI-empowered techniques.

Current network generations including 5G are based on the postulates of an uncontrollable propagation environment that can be equalized through the design of sophisticated transmission and reception schemes. To achieve the ultra-high throughput, low-latency and extremely-high reliability expected for B5G networks, optimizing the communication end-points, i.e., transmitters and receivers, might not be enough to deliver satisfactory results~\cite{DiRenzo2020}. Conversely, the emerging concept of Smart Radio Environment (SRE), i.e., a programmable radio environment wherein the signal propagation can be effectively controlled, may turn the channel into an optimization variable---to be jointly optimized with involved beamforming patterns of transmitters---and go beyond its long-standing perceived adversary nature.   

Interestingly, the constituting element of SREs are Reconfigurable Intelligent Surfaces (RISs), which in their most common implementation are essentially passive reflectarrays whose antenna elements can be electronically controlled to backscatter and phase-shift the electromagnetic waves impinging upon them. At a mature development stage, they are expected to be inexpensive enough to be shaped as thin composite material sheets covering portions of walls and buildings, just like conventional wallpapers.

Envisioned applications of RISs in SREs focus on improving wireless system performance in terms of maximum achievable rate. For instance, RISs can improve coverage in cellular dead-zones by effectively altering the reflected direction of impinging electromagnetic waves in order to avoid blocked LoS links. Moreover, such programmable beam steering capability can reduce the interference (and in some cases the ElectroMagnetic Field Exposure (EMFE)) with other directions by confining the power within a specific solid angle. RISs unveil unprecedented opportunities in this regard by overcoming the fundamental limitations of the classical wireless channel that was considered as a black-box: the installation of RISs can further improve the capacity of a wireless system by encoding and modulating additional information via a controlled sequence of reflections. 
Therefore, B5G network deployments will heavily rely on SREs built upon RISs, which might be owned by network operators and efficiently installed on best-suited buildings to increase the experienced Quality-of-Service (QoS) in target areas (e.g., densely populated urban areas). 

However, this might require an ad-hoc pre-deployment phase of such passive devices that, in turn, would have a significant impact on network-based localization performance. In general, optimal RIS deployments attaining diverse key performance indicators (KPIs) is an open research problem, which may leverage on general optimization frameworks, e.g., providing $k$-coverage planning solutions where each generic test point is covered by at least $k$ direct or reflected beams. We envision a fully-deployed ubiquitous RIS-empowered scenario to complement existing ground networks due to the RIS low-complexity and low-cost properties. While resilient initial deployments might be still impaired when a disaster strikes, such scenario would likely offer an additional level of reliability. 

Historically, localization solutions have relied only on the LoS link, corresponding to the direct path between transmitter and receiver, which could be easily related to the actual user position. Although modern techniques can overcome such dependence by capitalizing on measurements from non-line-of-sight (NLoS) links, e.g. through simultaneous localization and mapping (SLAM), the inherent electromagnetic interactions happening in the physical environment are uncontrollable, thus yielding the sub-optimality of the localization process. 
One obvious application of the RIS concept is to circumvent LoS blockage but more sophisticated approaches are possible. For instance, due to the wavefront curvature in the near-field of a large RIS, it is possible to determine unknown clock biases by combining Phase-of-Arrival (PoA) and Time-of-Arrival (ToA) information. Moreover, the PoA sensed by an antenna array may be directly used to perform spherical-wave localization. In harsh environments such as indoor industry 4.0 scenarios, RISs can maintain consistent multipath thereby allowing dynamically accounting for object movements~\cite{Wymeersch2020}.

\begin{figure}[t!]
      \centering
      \includegraphics[clip, trim= {0 0 0 0}, width=0.9\linewidth ]{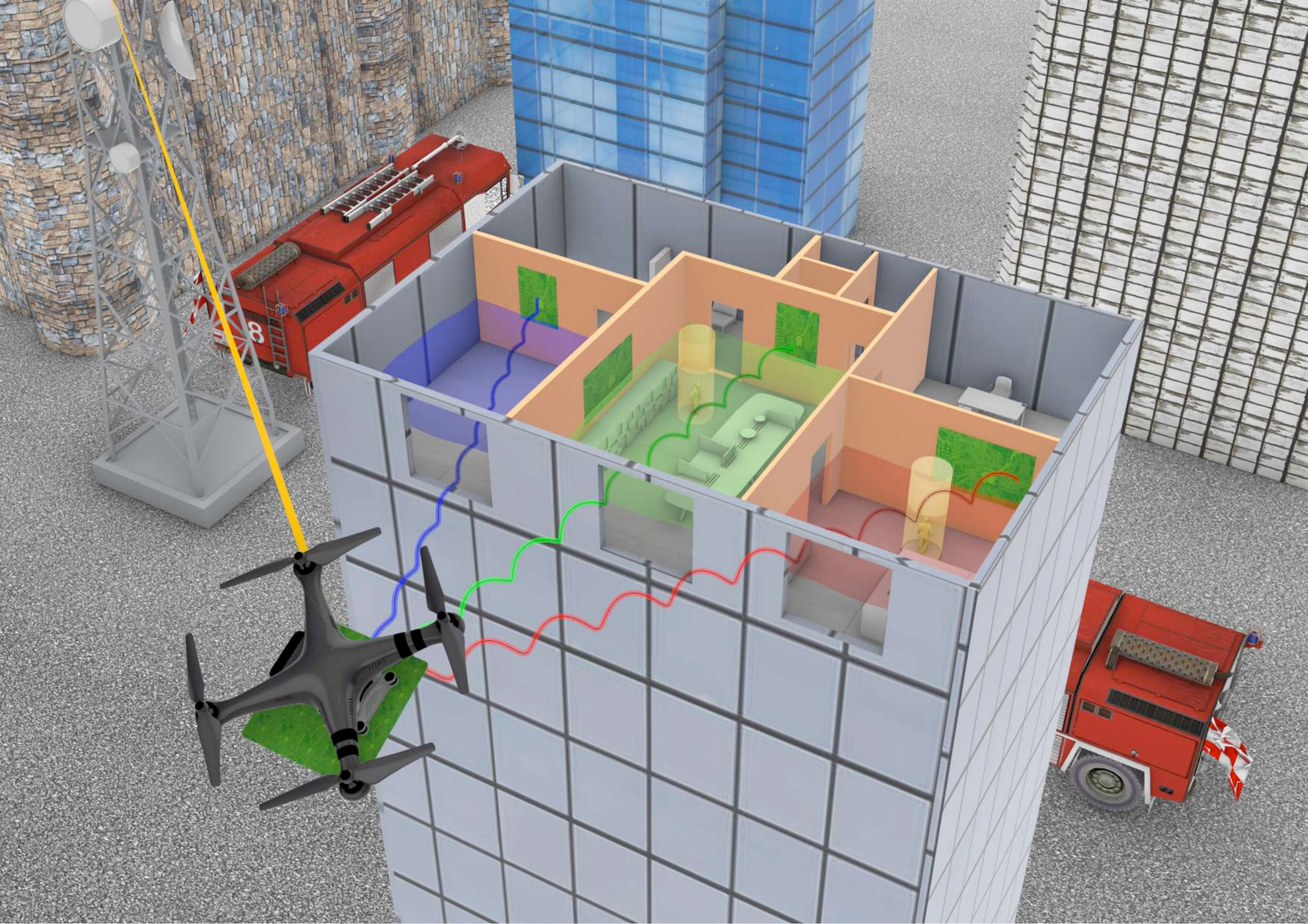}
      \caption{B5G Smart Radio Environment (SRE) design leveraging on Reconfigurable Intelligent Surfaces (RISs).}
      \label{fig:ris}
\end{figure}

Additionally, lightweight and low-complex RISs may be exploited to cope with the impelling energy-consumption issue of aerial devices to be used to bring connectivity capabilities to hard-to-reach locations. In particular, while the usage of UAVs---as proved in Section~\ref{s:pseudo-mult}---is effective to develop sophisticated localization solutions, it might be further improved when RISs are installed on-board, as shown in Fig.~\ref{fig:ris}. The RIS can be automatically controlled to focus the incoming signal towards specific locations while assisting indoor localization processes by means of RISs installed on the wall/window glass. This would offer additional means to first responder teams in emergency situations, when for e.g. the smog may impair the normal visibility inside the building.

\subsection{Seamless technology integration}
The above-mentioned enablers can be seamlessly integrated to deliver high localization accuracy in emergency scenarios. Indeed, pseudo-multilateration may benefit from the higher accuracy on single-link distance estimation enabled by OTFS modulation and RISs deployed in the environment. While OTFS may help to counteract the Doppler shift whenever high-speed devices are involved, RISs may be used as a low-cost alternative for a multi-antenna transmitter on-boarded on the UAV, as described in Section~\ref{s:ris}. Specifically, UAV-mounted RISs may focus the impinging signals from some survivor ground BSs or some portable emergency BSs towards specific locations by means of their passive beamforming capabilities and, in turn, improve the communication quality with the ground UEs. Besides, second-order reflections on RISs installed on buildings walls may further improve the performance. Due to the low-complexity and the low-cost properties of the novel RIS technology, RISs will likely be amost ubiquitous, thereby improving the reliability of the ground network infrastructure in emergency scenarios.

\section{Conclusions}
\label{s:conclusions}

UAVs are proving to be ideal candidates to enrich the cellular network technology in the B5G era. Thanks to their deployment flexibility, they allow accessing difficult-to-reach areas and enable the execution of unprecedented localization techniques, which can now take advantage of the anchor points mobility.  
In this paper, we have presented a novel localization solution, named pseudo-multilateration, that extends the concept of classical multilateration by assuming a single moving anchor, e.g., a flying UAV, taking distance measurements from a ground UE over time. We have tackled the localization problem in B5G from three different perspectives: $i$) \textit{coding}: a CNN derives the UE position by processing the measurements as a single-channel image and compensates for the channel shadowing induced by obstacles in the scenario; 
$ii$) \textit{modulation}: OTFS converts a time-variant OFDM channel into a non-fading time-invariant channel, yielding resilience against the Doppler spread induced by high-mobility conditions; 
$iii$) \textit{propagation}: SREs obtained by deploying RISs turn the channel into an optimization variable overcoming its uncontrollable nature.

\bibliographystyle{IEEEtran}
\bibliography{bibliography}

\begin{IEEEbiography}[{\includegraphics[width=1in,height=1.25in,clip,keepaspectratio]{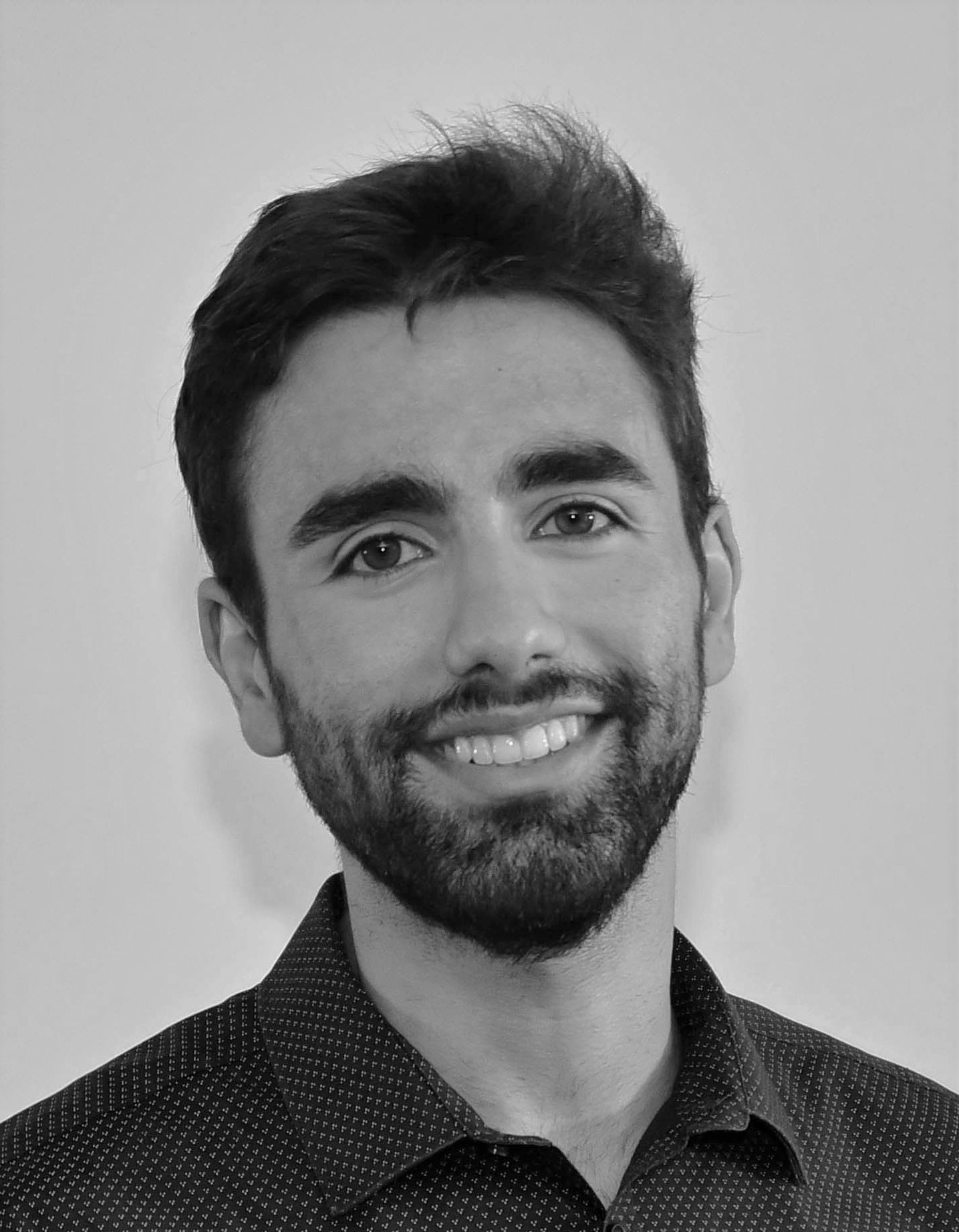}}]{Antonio Albanese} (S'18) received the M.Sc. in Telecommunications Engineering from Politecnico di Milano in 2018. Currently, he is pursuing his Ph.D. in Telematic Engineering at Universidad Carlos III de Madrid while being appointed as Research Associate at NEC Laboratories Europe GmbH. His research field covers optimization, machine learning techniques, blockchain and MEC topics with a particular interest in positioning and prototyping.
 \end{IEEEbiography}

 \begin{IEEEbiography}[{\includegraphics[width=1in,height=1.25in,clip,keepaspectratio]{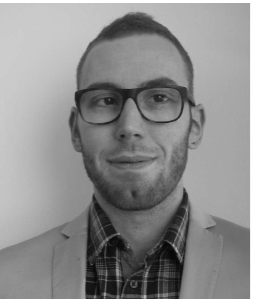}}]{Vincenzo Sciancalepore} (S'11--M'15--SM'19) received his M.Sc. degree in Telecommunications Engineering and Telematics Engineering in 2011 and 2012, respectively, whereas in 2015, he received a double Ph.D. degree. Currently, he is a principal researcher at NEC Laboratories Europe in Heidelberg, focusing his activity on network virtualization and network slicing challenges. He is the Chair of the ComSoc Emerging Technologies Initiative (ETI) on Reconfigurable Intelligent Surfaces (RIS) and an editor of IEEE Transactions on Wireless Communications.
 \end{IEEEbiography}

 \begin{IEEEbiography}[{\includegraphics[width=1in,height=1.25in,clip,keepaspectratio]{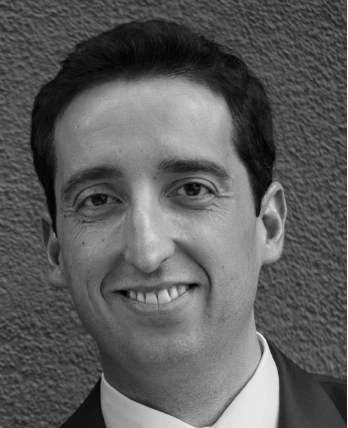}}]{Xavier Costa-P\'erez} (M'06--SM'18) is ICREA Research Professor, Scientific Director at the i2Cat Research Center and Head of 5G Networks R\&D at NEC Laboratories Europe. He has served on the Organizing Committees of several conferences, published papers of high impact and holds tenths of granted patents. Xavier received  his  Ph.D. degree in Telecommunications from the Polytechnic University of Catalonia (UPC) in Barcelona and was the recipient of a national award for his Ph.D. thesis.
 \end{IEEEbiography}

\end{document}